# Gradient-Boosted Pseudo-Weighting: Methods for Population Inference from Nonprobability samples

Kangrui Liu[1,^], Lingxiao Wang[2,^], Yan Li[1,3,*]

[1] Joint Program in Survey Methodology, University of Maryland, College Park, U.S.A
[2] Department of Statistics, University of Virginia, U.S.A
[3] Department of Epidemiology and Biostatistics, University of Maryland, College Park, U.S.A
[^] Two authors contributed equally
[*] Correspondence: Yan Li, Joint Program in Survey Methodology and Department of Epidemiology and Biostatistics, University of Maryland, College Park, MD, U.S.A. Email: yli6@umd.edu

## Abstract

Nonprobability samples have rapidly emerged to address time-sensitive priority topics in a variety of fields. While these data are timely, they are prone to selection bias. To mitigate selection bias, a large number of survey research literature has explored the use of propensity score (PS) adjustment methods to enhance population representativeness of nonprobability samples, using probability-based survey samples as external references. A recent advancement, the 2-step PS-based pseudo-weighting adjustment method ($_2$PS, Li 2024), has been shown to improve upon recent developments with respect to mean squared error. However, the effectiveness of these methods in reducing bias critically depends on the ability of the underlying propensity model to accurately reflect the true selection process, which is challenging with parametric regression. In this study, we propose a set of pseudo-weight construction methods, which utilize gradient boosting methods (GBM) to estimate PSs in $_2$PS to construct pseudo-weights, offering greater flexibility compared to logistic regression-based methods. We compare the proposed GBM-based pseudo-weights with existing methods, including $_2$PS. The population mean estimators are evaluated via Monte Carlo simulation studies. We also evaluated prevalence of various health outcomes, including 15-year mortality, using 1988 ~ 1994 NHANES III as a nonprobability sample and the 1994 NHIS as the reference survey.

**Keywords: Gradient Boosting, Non-probability Sample, Selection Bias, Propensity Scores, Population Representation**

# 1. Introduction

In the world of "big data" with fast collection of nonprobability samples, probability samples have been serving an important role as a reference. Various propensity score (PS)-based methods, which compare nonprobability samples to reference samples, have been proposed to reduce selection bias. These methods are generally grouped into two categories: 1) PS-weighting (PSW), where pseudo-weights are constructed by the inverse of the PS (Chen et al., 2020; Elliott, 2013; Valliant & Dever, 2011) or the odds of PS (Wang et al., 2021); and 2) PS-matching (PSM), where PS values or their monotone transformations are used as similarity measures to distribute survey weights to nonprobability sample units with similar PS measures (Lee & Valliant, 2009; Wang et al., 2020; Rivers, 2007). Some reviews on methods of statistical data integration for finite population inference can be consulted in Buelens et al. (2018), Valliant (2020), Yang and Kim (2020), Rao (2021), etc.

Both PSW and PSM require estimation of the PS under the assumption of conditional exchangeability (Li, 2024). Various PS estimation methods have been explored. One (unweighted) approach estimates PS's by comparing the non-probability sample with an *unweighted* probability sample, aiming to remove selection bias by balancing confounders (i.e., the covariates that are associated with both nonprobability participation and the outcome of interest) between the two samples, using either a parametric regression model (Wang et al., 2020; River, 2007) or nonparametrically via various machine learning methods (Kern et al., 2021). These methods, however, can yield biased population estimates if the distributions of confounders in the (unweighted) probability sample differ from those in the finite population (Wang et al., 2022). As a remedy, weighted PS estimation has been developed by comparing the non-probability sample with a *weighted* probability sample that can be representative of the target finite population (FP). The resulting pseudo-weighted estimators of the FP quantities, although are approximately unbiased when the propensity model is correctly specified (Chen et al., 2020; Wang et al., 2021), can be statistically inefficient. This is because the differential weights are highly variable among the large sample weights in the probability survey and the unit weights (i.e., weight =1) for individuals in the nonprobability sample, and therefore can introduce substantial variance in the PS estimation.

To achieve efficiency while maintaining unbiasedness, Li (2024) proposed a 2-step PS estimation method ($_2$PS). In the first step, the PS is estimated using an unweighing approach, which may lead to biased estimates of the FP quantities; In the second step, the potential bias is corrected by adjusting for confounders whose distributions in the (unweighted) probability sample differ from those in the FP. The procedure estimates the PS using traditional parametric techniques (e.g., logistic regression), motivated by their simplicity and interpretability. However, in the context of nonprobability sample inference, the PS is not estimated for interpretive purposes. Instead, it is used to construct pseudo-weights for the nonprobability sample, with the goal of reducing selection bias and achieving balance in the distribution of confounders between the pseudo-weighted nonprobability sample and the sample-weighted probability sample. Furthermore, the validity of these estimates is based on parametric models, and critically depends on correctly specifying the functional form of the predictors, including all relevant interactions and nonlinear effects. Model misspecification can lead to poorly balanced covariate distributions, producing biased estimates (Lee & Little, 2017; Salditt & Nestler,

2023).

Nonparametric machine learning (ML) methods (e.g., boosted regression trees, Bayesian additive regression trees, super learner, etc.) have been shown to flexibly capture complex interactions and nonlinearities, and typically achieve superior empirical covariate balancing and lower bias in the presence of model uncertainty (Hill et al, 2020; Pirracchio et al 2015; Castro et al., 2020; Chu & Beaumont, 2019; Ferri-García & Rueda, 2020). This flexibility enhances covariate balance in practice, particularly in settings with high-dimensional covariate spaces or complex multivariate relationships (Wyss et al, 2014). Incorporating machine learning algorithms into the PS estimation facilitates a reliable balance between extensive sets of predictors (Hejazi & Laan, 2022).

Recently, ML algorithms have been considered in the context of improving population representativeness of nonprobability samples (see, e.g., Mercer et al., 2018; Ferri-García & Rueda, 2020; Buelens et al., 2018; Kern et al., 2021; Chu & Beaumont, 2019; Castro et al., 2021; Liu et al., 2024). Among the most employed ML methods for PS estimation, gradient boosting algorithms (Leite et al., 2024) are shown, on average, to yield better results than other selected ML methods (Kern et al. 2021; Rueda et al. 2022; Rueda et al. 2024). Also, it is supported by a robust user community in R program language. The *twang* package (Ridgeway et al., 2013) offers a dedicated implementation with built-in covariate balance diagnostics and data-driven stopping rules. We propose the use of the gradient boosting method for estimating propensities to construct pseudo-weights for selection bias reduction.

Building upon the theoretical framework established by Li (2024), in this paper we propose a boosted 2-step procedure (**Boost**$_2$PS), which introduces gradient boosting methods to estimate propensity scores, improving estimation accuracy in the presence of complex covariate interactions and nonlinearities. Through our simulation studies and real-world applications, **Boost**$_2$PS consistently outperforms the original $_2$PS method, especially in moderate to severe nonlinearity scenarios. Recognizing that $_2$PS extends the single-stage weighted estimator ($_1$PS; Wang et al., 2021), we also propose a boosted version of $_1$PS, denoted as **Boost**$_1$PS, to enable a comparative evaluation with $_2$PS and **Boost**$_2$PS.

The article is structured as follows. Section 2 describes the framework of FP inferences using nonprobability samples with a brief review of the logistic regression-based PS methods. Section 3 introduces the proposed GBM-based PS adjustment methods, detailing the hyperparameter optimization process. Section 4 evaluates the performance of the proposed methods compared to the logistic regression-based PS methods under various scenarios with different levels of complexity of the nonprobability sample participation, The proposed methods are applied in a real-world scenario in Section 5. Finally, the implications of our findings are discussed in Section 6.

## 2. Basic Setting and Existing Methods

### 2.1 Basic setting

We are interest in estimating the mean $\mu$ of a variable $Y$ in a target finite population ($FP$) of $N$ individuals:

$$\mu = N^{-1} \sum_{i \in U} y_i,$$

where $U = \{1, \cdots, N\}$ denotes the set of all $FP$ individuals and $\{y_1, \cdots, y_N, i \in U\}$ is the realization of $Y$ in the $FP$. Suppose $s_c \subset U$ is a volunteer-based nonprobability sample of size $n_c$ recruited from the $FP$ by a self-selection participation mechanism, with $\delta_i^{(c)}(=1$ if $i \in s_c; 0$ otherwise), denoting the participation of $s_c$ or not. The underlying probability that an $FP$ unit $i$ is self-selected into the nonprobability sample (i.e., participation rate) is:

$$\pi_i^{(c)} = P(i \in s_c|U) = E_c\left\{\delta_i^{(c)}\middle|y_i, \boldsymbol{x}_i\right\}, i \in U,$$

where the expectation $E_c$ is with respect to the nonprobability sample participation, and $\boldsymbol{x}_i$ is a vector of participation variables, that is, the covariates related to the probability of participating in $s_c$. The corresponding implicit nonprobability sample weights are $\left\{w_i = 1/\pi_i^{(c)}, i \in s_c\right\}$.

Ignoring the unequal participation rates $\left\{\pi_i^{(c)}, i \in U\right\}$ can lead to selection bias in estimating $\mu$ when the participation mechanism for $s_c$ (i.e., $\delta_i^{(c)}$) is associated with $y$, such that $E(y|s_c) \neq E(y|U)$. In order to reduce the selection bias, we consider the following regularity assumptions for the nonprobability sample participation analogous to Wang et al. (2020).

**A1.** The nonprobability sample participation is uncorrelated with the variable of interest given the observed covariates, which is $\pi_i^{(c)} = E_c\left\{\delta_i^{(c)}\middle|y_i, \boldsymbol{x}_i\right\} = E_c\left\{\delta_i^{(c)}\middle|\boldsymbol{x}_i\right\}, i \in U$.

**A2.** All $FP$ individuals have positive probabilities to be observed in the nonprobability sample, that is, $\pi_i^{(c)} > 0, i \in U$.

**A3.** The indicators of participation in the nonprobability are uncorrelated with each other given the observed covariate, that is, $cov\left(\delta_i^{(c)}, \delta_j^{(c)}\middle|\boldsymbol{x}_i, \boldsymbol{x}_j\right) = 0, i \neq j, i, j \in U$.

The assumption **A1** implies conditional exchangeability, i.e., equality of conditional expectations, $E(y|b(\boldsymbol{x}), s_c) = E(y|b(\boldsymbol{x}), U)$, where $b(\boldsymbol{x})$ is a balancing score defined as

a function of the observed covariates $x$ (Li, 2024). The positivity assumption **A2** is required to ensure the estimation and identification of conditional expectations across the full support of covariates. **A3** implies that selection into the nonprobability sample is conditionally independent across units, thereby allowing standard variance estimation procedures for nonprobability sample inference to remain valid.

## 2.2 Existing Parametric Propensity Score Weighting Methods Using an Adaptive Balancing Score

Li (2024) proposed a two-step propensity weighting approach that estimates the implicit nonprobability sample participation weights $\{w_i, i \in s_c\}$ from an adaptive balancing score $b(x)$ by using a probability-based survey sample $s_s$ as the reference. The survey sample $s_s \subset U$ is randomly selected from the same target $FP$ as the nonprobability sample $s_c$, and has the sample weights $\{d_i, i \in s_s\}$. We assume that we can observe the nonprobability sample participation variables $\{x_i, i \in s_s\}$ in $s_s$, but not the outcome of interest $Y$.

In **Step 1**, a logistic regression model is fitted to the combined sample $s_c + s_s$ (regardless of whether units overlap) without considering the survey sample weights $\{d_i, i \in s_s\}$, which is given by

$$\log \frac{p_i}{1 - p_i} = \boldsymbol{\beta}^\top g(x_i), i \in s_c + s_s, \tag{2.1}$$

where $p_i$ is the propensity of being in $s_c$ vs. in $s_s$ for individual $i \in s_c + s_s$ and $b_1(x; \boldsymbol{\beta}) = \boldsymbol{\beta}^\top g(x)$ balances the distribution of $x$ in unweighted $s_c$ vs. unweighted $s_s$. We obtain the estimates of $b_1(x; \boldsymbol{\beta})$ and $p_i$, respectively, denoted by $b_1(x; \widehat{\boldsymbol{\beta}})$ and $\hat{p}_i =$ $\text{expit}\{b_1(\boldsymbol{x}_i; \widehat{\boldsymbol{\beta}})\}$, where $\widehat{\boldsymbol{\beta}}$ are the estimates of the parameters $\boldsymbol{\beta}$ in model (2.1).

In **Step 2**, a logistic regression model is fitted to the combined set $s_s + U$, which is approximated by the combined unweighted $s_s$ and weighted $s_s$ as follows

$$\log \frac{q_i}{1 - q_i} = \boldsymbol{\gamma}^\top g(\boldsymbol{x}_i), i \in s_s, \tag{2.2}$$

where $b_2(\boldsymbol{x}; \boldsymbol{\gamma}) = \boldsymbol{\gamma}^\top g(\boldsymbol{x}_i)$ balances the distribution of $\boldsymbol{x}$ in unweighted $s_s$ vs. $U$, represented by the weighted $s_s$, and $q_i = \Pr(i \in s_s | s_s + U)$. The estimates of $\boldsymbol{\gamma}$, $b_2(\boldsymbol{x}; \boldsymbol{\gamma})$, and $q_i$ are denoted by $\widehat{\boldsymbol{\gamma}}$, $b_2(\boldsymbol{x}; \widehat{\boldsymbol{\gamma}})$, and $\widehat{q}_i$, respectively.

The final balancing score $b(\boldsymbol{x}; \boldsymbol{\beta}, \boldsymbol{\gamma}) = (\boldsymbol{\beta} + \boldsymbol{\gamma})^\top g(\boldsymbol{x}_i)$ is then constructed to balance the distribution of $\boldsymbol{x}$ in the naïve $s_c$ and that in $U$ (represented by the weighted $s_s$). The final set of pseudo-weights for the nonprobability sample $s_c$ can be calculated by the 2-step PS-based method (₂PS): $\widehat{w}_i^{2\text{PS}} = \exp\{-b(\boldsymbol{x}_i; \widehat{\boldsymbol{\beta}}, \widehat{\boldsymbol{\gamma}})\}, i \in s_c$.

The ₂PS method is shown to be more efficient than the one-step adaptive logistic propensity (₁PS) weighting method which estimates the logistic regression model (2.1) using the combined data from the nonprobability sample and the *weighted* probability survey sample. The ₁PS pseudo-weights $\widehat{w}_i^{1PS} = \exp\{-\widehat{\boldsymbol{\beta}}_w^\top g(\boldsymbol{x}_i)\}$, $i \in s_c$ with $\widehat{\boldsymbol{\beta}}_w$ being the estimates of $\boldsymbol{\beta}$ obtained from the weighted sample (see more details in Section 3.5).

However, the ₂PS method requires fitting two parametric PS models in the first and second steps, both assumed to have logit links and share the same covariate function $g(\boldsymbol{x})$. Misspecification in either step can limit the reduction of selection bias. Selecting variables for $g(\boldsymbol{x})$ and assessing the goodness-of-fit of the logistic propensity model in the second step are relatively straightforward. *First*, survey design variables that should be included in $g(\boldsymbol{x})$ are often known from the documentation of well-designed probability-based reference surveys. *Second*, both the functional form of $g(\boldsymbol{x})$ and the model fit can be evaluated by minimizing the difference between the known true survey weights $\{d_i, i \in s_s\}$ and their estimated counterparts $\{\hat{d}_i = \hat{q}(\boldsymbol{x}_i)/(1 - \hat{q}(\boldsymbol{x}_i)), i \in s_s\}$. Assessing the model goodness-of-fit in Step 1, however, can be more challenging. It is often of a question, e.g., whether using a linear combination $b_1(\boldsymbol{x}) = \boldsymbol{\beta}^\top g(\boldsymbol{x}_i)$ in the first step with the same functional form $g(\boldsymbol{x})$ as that in the second step is adequate. As a result, adopting a more flexible functional form for $b_1(\boldsymbol{x})$ in the first step is crucial for improving the accuracy of estimating the nonprobability sample weights $\{w_i, i \in s_c\}$.

## 3. Proposed Boosted Two-Step Propensity Weighting Method

We propose to enhance the original $_2$PS weighting approach by combining a flexible machine learning approach and a logistic regression model. In the first step, we estimate the balancing score $b_1(\boldsymbol{x})$ that balances the distribution of $\boldsymbol{x}$ in $s_c$ vs. unweighted $s_s$ using GBM. Then, we combine the GBM balancing score calculated in the first step and the balancing score $b_2(\boldsymbol{x}, \hat{\boldsymbol{\gamma}})$ estimated from logistic propensity model (2.2) in the second step to obtain the final pseudo-weights.

We expect that GBM performs well in analyzing potentially complex functional forms, while the degree of complexity can be precisely regulated through their associated hyperparameters. In the next section, we describe the details of how the GBM balancing score and the final pseudo-weights are calculated.

### 3.1 Gradient Boosting Method for Balancing Score Estimation

Gradient Boosting Machine (GBM) is an ensemble method that iteratively forms and sums up a group of simple regression tree models to minimize prediction error or the loss function (Friedman 2001). It is considered as a powerful tool to estimate the propensity of treatment assignment in causal inference framework (McCaffrey et al., 2004; Lee et al., 2010). GBM distinguishes itself from the traditional boosting methods by using gradient descent rather than reweighting the misclassified data to minimize the loss function. Gradient descent directly minimizes the loss function through iterative parameter updates, ensuring stable convergence, while reweighting misclassified data often relies on manually tuned sample weights and may introduce instabilities (An et al., 2020). Therefore, it is naturally to consider using GBM to estimate the propensity of being in the $s_c$ vs. in the $s_s$ in our setting. Furthermore, GBM, different from a lot of other machine learning methods which directly models the propensity scores, models the log-odds of the propensity, i.e., the balancing score $b_1(x)$ (McCaffrey et al., 2004), and therefore better fits the two-step PS weighting approach that combines two balancing scores for the final propensity estimation.

Unlike the logistic propensity model (2.1) which assumes a linear relationship between the balancing score and $g(x)$, i.e., $b_1(x; \beta) = \beta^\top g(x)$, GBM allows for a more complex and flexible tree-based form of $b_1(x)$, obtained by minimizing the log-loss function (3.1) iteratively (McCaffrey et al., 2004), given by

$$l(b_1) = \sum_{i \in s_c + s_s} (R_i b_1(x_i) - \log[1 + \exp\{b_1(x_i)\}]),$$

(3.1)

where $R_i$ is the indicator of sample membership ($= 1$ if $i \in s_c$, and $= 0$ if $i \in s_s$). The initial value of $b_1(x)$ is the non-parametric sample log-odds of the propensity of being in $s_c$ vs. $s_s$, that is $b_1^{(0)}(x) = \log(n_c/n_s)$. In the $t$-th iteration, GBM first calculates the pseudo residuals of the $(t-1)$-th model, denoted by $r_i^{(t-1)}$ for $i \in s_c + s_s$ as follows:

$$r_i^{(t-1)} = \frac{\partial l(b_1)}{\partial b_1(x_i)} \bigg|_{b_1(x_i) = b_1^{(t-1)}(x_i)} = R_i - \hat{p}^{(t-1)}(x_i), \qquad i \in s_c + s_s,$$

where $\hat{p}^{(t-1)}(\boldsymbol{x}_i) = \text{expit}\left\{b_1^{(t-1)}(\boldsymbol{x}_i)\right\}$ is the propensity of being in $s_c$ vs. $s_s$, for $i \in s_c + s_s$, estimated from the $(t-1)$-th model, with the initial value $p^{(0)} = n_c/(n_c + n_s)$.

Then, the pseudo residuals $\left\{r_i^{(t-1)}, i \in s_c + s_s\right\}$ are used as the values of the "response variable" to fit a new weak learner and obtain a decision tree for residuals below

$$h^{(t)}(\boldsymbol{x}) = \sum_{m=1}^{M_t} \alpha_m^{(t)} \text{I}\left(\boldsymbol{x} \in \tau_m^{(t)}\right),$$

where $M_t$ is the total number of terminal nodes of $t$-th tree , $\text{I}\left(\boldsymbol{x} \in \tau_m^{(t)}\right)$ is an indicator function, indicating whether the input $\boldsymbol{x}$ falls into the $m$-th terminal node $\tau_m^{(t)}$, and $\alpha_m^{(t)}$ is the mean of residuals, $\left\{r_i^{(t-1)}, \boldsymbol{x}_i \in \tau_m^{(t)}\right\}$, of the $t$-th tree at the $m$-th terminal node, that is, the average of residuals falling into the $m$-th node.

Then $t$-th model combines the $(t-1)$-th model and the new decision tree $h^{(t)}(\boldsymbol{x})$ for the residuals by the shrinkage parameter $\nu$:

$$b_1^{(t)}(\boldsymbol{x}) = b_1^{(t-1)}(\boldsymbol{x}) + \nu \cdot h^{(t)}(\boldsymbol{x}).$$

The shrinkage parameter $\nu$ controls the contribution of the new tree $h^{(t)}(\boldsymbol{x})$ to the final model to avoid overfitting. Eventually, the final GBM model is the weighted sum of all weak learners as follows:

$$b_1^{(T)}(\boldsymbol{x}; \boldsymbol{\theta}) = b_1^{(0)}(\boldsymbol{x}) + \nu \sum_{t=1}^{T} h^{(t)}(\boldsymbol{x}),$$

where $\boldsymbol{\theta} = (\nu, T, M)^{\top}$ is a vector of tunning parameters including the shrinkage parameter $\nu$, number of trees $T$, and the maximum depth of each tree $M$ (to decide the maximum number of terminal nodes).

## 3.2 Final Adaptive Balancing Score and Pseudo Sample Weighted Mean

We create the final Boost$_2$PS that balances the distribution of $\boldsymbol{x}$ in the naïve $s_c$ and that in $U$ by combining the GBM-based $\hat{b}_1^{(T)}(\boldsymbol{x}; \boldsymbol{\theta})$ in the first step and the logistic model-based $b_2(\boldsymbol{x}, \hat{\boldsymbol{\gamma}})$ in the second step described in Section 2 as $b(\boldsymbol{x}_i; \boldsymbol{\theta}, \hat{\boldsymbol{\gamma}}) = \hat{b}_1^{(T)}(\boldsymbol{x}_i; \boldsymbol{\theta}) + b_2(\boldsymbol{x}_i, \hat{\boldsymbol{\gamma}})$, for $i \in s_c$ and construct the pseudo-weights $\left\{ \hat{w}_i^{Boost_2PS} = \exp\left\{ -\hat{b}_1^{(T)}(\boldsymbol{x}_i, \boldsymbol{\theta}) - b_2(\boldsymbol{x}_i, \hat{\boldsymbol{\gamma}}) \right\}, \ i \in s_c \right\}$. Finally, we estimate the $FP$ quantity $\mu$ from the pseudo-weighted $s_c$ as follows.

$$\hat{\mu}^{Boost_2PS} = \frac{\sum_{i \in s_c} \hat{w}_i^{Boost_2PS} y_i}{\sum_{i \in s_c} \hat{w}_i^{Boost_2PS}}, \tag{3.2}$$

## 3.3 Tuning Hyper-Parameters for GBM Balancing Score Estimation

As a machine learning approach, GBM requires the specification of tuning parameters before the model can be built. As described in Section 3.1, $\boldsymbol{\theta} = (\nu, T, M)^\top$ are crucial tunning parameters and can affect the performance of the GBM in estimating the balancing score.

As we aim to balance the distribution of $\boldsymbol{x}$ in the $s_c$ and unweighted $s_s$ via the GBM balancing score adjustment in Step 1, we choose the combination of the tunning parameters $\boldsymbol{\theta} = (\nu, T, M)^\top$ from a set of pre-specified candidate values that minimizes the difference between the distributions of $\boldsymbol{x}$ in the two samples measured by the averaged absolute standardized mean differences (ASMD), defined as follows.

$$\text{ASMD} = \sum_{j=1}^{J} a_j \frac{\left| \bar{x}_c^{(j)} - \bar{x}_s^{(j)} \right|}{\sqrt{\frac{1}{2}\left(\sigma_c^{2(j)} + \sigma_s^{2(j)}\right)}}, \tag{3.3}$$

where $\bar{x}_c^{(j)}$ and $\sigma_c^{2(j)}$ are the sample mean and variance of the $j$-th covariate in $s_c$ with the GBM weights $\left\{ \widehat{w}_i^{GBM} = \exp\left\{ -\hat{b}_1^{(T)}(\boldsymbol{x}_i, \boldsymbol{\theta}) \right\}, i \in s_c \right\}$; $\bar{x}_s^{(j)}$ and $\sigma_s^{2(j)}$ are the sample mean and variance of the $j$-th covariate in the unweighted $s_s$; $\{a_j, j = 1, \cdots, J\}$ is a set of constants measuring the contribution of the $j$-th covariate to the ASMD. For example, if $a_1 = \cdots = a_J = 1$, all covariates are considered equally important. If $a_j$ are set to be the absolute values of coefficients of the regression outcome model for $Y$, the covariates that are more predictive of the outcome are considered to be more important.

The pre-specified candidate values for tunning parameters $\boldsymbol{\theta} = (\nu, T, M)^\top$ can vary depending on factors such as the sample sizes of $s_c$, and $s_s$, the number of the considered covariates, and the complexity of data structures, etc. Following Friedman et al. (2001), we consider small values of 0.1, 0.01, and 0.001 for shrinkage parameter $\nu$ to achieve better generalization performance. The number of trees should be large when using a small $\nu$, typically ranging from 1000 to 5000. We consider tree depths of 2, 3, 4, and 5, starting with shallow trees for simple functional forms and increasing the depth as the data structure becomes more complex. More details are described in Section 4.

## 3.4 Variance Estimation

We use bootstrap replication approach to estimate the variance of the proposed estimator $\hat{\mu}^{Boost_2PS}$. For the $l$-th bootstrap replicate, we resample $n_c$ individuals from the original nonprobability sample $s_c$ using simple random sampling with replacement (SRSWR). For $s_s$, we consider a general case where there are $H$ strata with $a_h$ primary sampling units (PSU's) in stratum $h$ in $s_s$, $h = 1, \cdots, H$. We resample $a_h - 1$ PSU's within stratum $h$ by SRSWR, for each unit $i$ in PSU $j$ within stratum $h$, we define the bootstrap weights $d_{hji}^{(l)}$

$$d_{hji}^{(l)} = \frac{a_h}{a_h - 1} m_{hj}^{(l)} d_i$$

where $m_{hj}^{(l)}$ is the number of times PSU $j$ is selected in the $l$-th bootstrap sample, and $d_i$ is the original survey sample weight of the survey sample unit $i$ in PSU $j$ within stratum $h$. Then we apply the proposed two-step GBM weighting approach to estimate the $FP$ parameter $\mu$ using the $l$-th bootstrap replicate sample with the bootstrap weights. The resulting estimate is denoted by $\hat{\mu}_l^{Boost_2PS}$. The bootstrap variance of the full sample estimate $\hat{\mu}^{Boost_2PS}$ is obtained by

$$\text{var}(\hat{\mu}^{Boost_2PS}) = \frac{1}{L-1} \sum_{l=1}^{L} (\hat{\mu}_l^{Boost_2PS} - \hat{\mu}^{Boost_2PS})^2,$$

where $L$ is the total number of bootstrap replicates.

## 3.5 Using GBM to Improve the Adaptive Logistic Propensity Weighting Approach

For comparison purpose, we also consider using GBM to improve the parametric $_1$PS approach proposed by Wang et al. (2021). The original $_1$PS method fits the logistic regression model (2.1) to the combined naïve $s_c$ and *weighted* $s_s$, and constructs pseudo-weights $\{\hat{w}_i^{ALP} = \exp\{-\hat{\boldsymbol{\beta}}_w^\top g(\boldsymbol{x}_i)\}, i \in s_c\}$ with $\hat{\boldsymbol{\beta}}_w$ being the estimates of $\boldsymbol{\beta}$. In order to consider the survey sample weights in GBM approach, consider the weighted loss function:

$$l_w(b_w) = \sum_{i \in s_c + s_s} d_i(R_i b_w(\boldsymbol{x}_i) - \log[1 + \exp\{b_w(\boldsymbol{x}_i)\}]), \quad (3.4)$$

where $d_i$ is the survey sample weight for $i \in s_s$ and $d_i = 1$ for $i \in s_c$, and the initial value for $b_w(\boldsymbol{x}_i)$ is $b_w^{(0)}(\boldsymbol{x}) = \log\{n_c / (\sum_{i \in s_s} d_i)\}$. The new weak learner below at $t$-th iteration is $h_w^{(t)}(\boldsymbol{x}) = \sum_{m=1}^{M_t} \alpha_{w,m}^{(t)} \, \mathrm{I}\left(\boldsymbol{x} \in \tau_m^{(t)}\right)$, with $\alpha_{w,m}^{(t)}$ being the weighted average of

$\left\{r_i^{(t-1)}, \boldsymbol{x}_i \in \tau_m^{(t)}\right\}$ for the $m$-th node, where $r_i^{(t-1)} = \partial l_w(b_w)/\partial b_w(\boldsymbol{x}_i)\big|_{b_w(\boldsymbol{x}_i)=b_w^{(t-1)}(\boldsymbol{x}_i)}$.

The tuning parameters are determined in a similar way with Section 3.3, but with the

ASMD calculated from the Boost₁PS-weighted $s_c$ *vs.* sample-weighted $s_s$. We

respectively denote the Boost₁PS pseudo-weights and resulting estimator of $\mu$ as

$\left\{\widehat{w}_i^{Boost_1PS}, \ i \in s_c\right\}$ and $\hat{\mu}^{Boost_1PS}$, which uses $\widehat{w}_i^{Boost_1PS}$ to replace $\widehat{w}_i^{Boost_2PS}$ in formula

(3.4).

## 3.6  R implementation of $\hat{\mu}^{Boost_2PS}$ and $\hat{\mu}^{Boost_1PS}$

We use the R function *gbm*() in the *gbm* package (Ridgeway, 2013) to fit the trees in the combined sample $s_c + s_s$ for the first step of the Boost₂PS method. It returns the balancing score $\hat{b}_1^{(T)}(\boldsymbol{x}_i, \boldsymbol{\theta})$ (i.e., the propensity of being in $s_c$ vs. $s_s$ on the logit scale),

which can be directly used to calculate the GBM weights $\Big\{\widehat{w}_i^{GBM} =$

$\exp\left(-\hat{b}_1^{(T)}(\boldsymbol{x}_i, \boldsymbol{\theta})\right), i \in s_c\Big\}$ and the final pseudo-weights $\Big\{\widehat{w}_i^{Boost_2PS} =$

$\exp\left\{-\hat{b}_1^{(T)}(\boldsymbol{x}_i, \boldsymbol{\theta}) - b_2(\boldsymbol{x}_i, \hat{\boldsymbol{\gamma}})\right\}, i \in s_c\Big\}$. To determine the optimal tuning parameters $\boldsymbol{\theta}$,

we use the R function *ps*() in the *twang* package (Ridgeway et al., 2022). It serves as an

interface to the *gbm*() function to compute ASMDs between the GBM weighted $s_c$ and

the unweighted $s_s$. Note that both *gbm*() and *ps*() provide the option of incorporating the

survey sample weights respectively in fitting the trees and calculating the ASMDs for the

Boost₁PS method. The point estimates $\hat{\mu}^{Boost_2PS}$ and $\hat{\mu}^{Boost_1PS}$ can be obtained using the R function *svymean*() in the *survey* package (Lumley, 2004), with the *weight* option set to be $\left\{ \widehat{w}_i^{Boost_2 PS}, i \in s_c \right\}$ or $\left\{ \widehat{w}_i^{Boost_1 PS}, i \in s_c \right\}$ respectively.

## 4. Simulations

### 4.1 Population Generation

In the $FP$ of size $N = 50,000$, we generated a vector of covariate $(X_1, \cdots, X_7)$ in two steps. First, 10 base covariates $(V_1, \cdots, V_7)$ were generated independently following standard normal distributions. Second, covariates $(X_1, \cdots, X_7)$ were generated using linear combinations of the base covariates with correlations, where $X_k = V_k$ for $k = 1, \cdots, 4, 7$, $X_5 = 1.171(0.16 V_1 + 0.84 V_5)$, and $X_6 = 1.353(0.67 V_2 + 0.33 V_6)$. The outcome variable of interest $Y$ was a binary variable generated by a Bernoulli distribution with mean expit($\boldsymbol{\alpha}^\top \boldsymbol{X}$), where $\boldsymbol{X} = (1, X_1, X_2, X_3, X_4, X_5, X_6, X_7)^\top$ and $\boldsymbol{\alpha} = (-2.5, 1, 1, 1, 1, 0, 0, 0)^\top$. We use $y_i$ to denote the value of $Y$ for the $FP$ unit $i$. The parameter of interest is the population mean of $Y$, that is, $\mu = N^{-1} \sum_{i \in U} y_i \approx 0.17$ in the simulation.

### 4.2 Sampling from the Finite Population to Assemble the Survey Sample and the Nonprobability Sample

The probability-based survey sample ($s_s$) and the nonprobability samples ($s_c$) were independently randomly selected from the $FP$ by probability proportional to size (PPS) sampling designs, with equal sample size of $n_s = n_c = 1,500$ but different measure of sizes (MOS). The MOS for selecting $s_s$ and $s_c$ are, respectively, $MOS_i^{(s)} = \exp(\boldsymbol{\beta}^\top \boldsymbol{x}_i)$ and $MOS_i^{(c)} = \exp(\boldsymbol{\eta}^\top \boldsymbol{u}_i)$ for the $FP$ unit $i \in U$, where $\boldsymbol{x}_i$ is the value of $\boldsymbol{X}$, $\boldsymbol{u}_i$ includes $\boldsymbol{x}_i$ and their functional forms (e.g. nonlinear higher order terms and non-additive interactions); $\boldsymbol{\beta}$ and $\boldsymbol{\eta}$ are the corresponding coefficients. We set $\boldsymbol{\beta} =$ $(0, 0.2, 0.2, 0.3, 0.3, -0.16, -0.1, 0.14)^\top$, with the first component corresponding to the intercept term and the remaining coefficients determining the strength of association between the covariates and the inclusion probability for sampling $s_s$. The value of $\boldsymbol{\eta}$ reflects the complexity of the relationship between $\boldsymbol{u}_i$ and nonprobability sample participation propensity $\pi^{(c)}$.

Under the PPS sampling designs, the survey sample selection probability $\pi_i^{(s)}$, i.e., the propensity of being in $s_s$ *vs.* $U$ can be formulated by:

$$\log\left(\pi_i^{(s)}\right) = \log\left(\frac{n_s MOS_i^{(s)}}{\sum_{i \in U} MOS_i^{(s)}}\right) = const_s + \boldsymbol{\beta}^\top \boldsymbol{x}_i,$$

where $const_s = \log n_s - \log \sum_{i \in U} MOS_i^{(s)}$. Similarly, the propensity of being in $s_c$ *vs.* $U$ can be formulated by

$$\log\left(\pi_i^{(c)}\right) = \log\left(\frac{n_c MOS_i^{(c)}}{\sum_{i \in U} MOS_i^{(c)}}\right) = const_c + \boldsymbol{\eta}^\top \boldsymbol{u}_i,$$

where $const_c = \log n_c - \log \sum_{i \in U} MOS_i^{(c)}$. As results, the propensity of being in $s_c$ *vs.* $s_s$ is given by:

$$\log\left(\frac{\pi_i^{(c)}}{\pi_i^{(s)}}\right) = \log\left(\frac{n_c MOS_i^{(c)}}{n_s MOS_i^{(s)}}\right) = const_c - const_s + \boldsymbol{\gamma}^\top \boldsymbol{u}_i,$$

where $\boldsymbol{\gamma} = \boldsymbol{\eta} - (\boldsymbol{\beta}^\top, \boldsymbol{0}^\top)^\top$. Therefore, the true underlying nonprobability sample participation propensities can be approximately unbiasedly estimated by the 1PS and the 2PS method, provided that all relevant covariates and their functional forms (in $\boldsymbol{u}_i$) are correctly specified in the logistic regression models. However, in practice, the true propensity models are unknown. To address this, we compare the performance of the proposed **Boost₂PS** and **Boost₁PS** method for estimating $\mu$, against the 1PS and 2PS methods based on logistic regression models that include the main effects of $X_1, \cdots, X_7$ and all two-way interactions for propensity estimation (Hirano and Imbens 2001; Dehejia and Wahba 2002).

We consider eight simulation scenarios with varied values of $\boldsymbol{\eta}$ in $MOS_i^{(c)}$ calculation (see Appendix A), which reflect the complexity of the relationship between covariates and their functional forms (in $\boldsymbol{u}_i$), and the nonprobability sample participation indictor $\delta$. As shown in the Figure 1, each scenario specifies a distinct functional form. All scenarios include main effects from $X_1, \cdots, X_7$. Scenario 1 is linear with no interaction or quadratic terms (denoted as $I_0Q_0$). Scenarios 2–4 ($I_0Q_1$, $I_1Q_0$, $I_1Q_1$) represent slightly nonlinear ($Q_1$) and/or nonadditive ($I_1$) structures. Scenarios 5–7 ($I_0Q_2$, $I_2Q_0$, $I_2Q_2$) are moderately nonlinear ($Q_2$) and/or nonadditive ($I_2$), while Scenario 8 ($I_3Q_3$) is severely nonlinear and nonadditive.

We evaluated the performance of the four methods using the following criteria. Relative Bias (RB%) was calculated as the bias—defined as the difference between the average of simulated estimates and the true population mean—divided by the population mean and multiplied by 100%. Empirical Variance was computed as the average of the squared deviations of the estimated means from their overall average across all replications. Variance Ratio (VR) was defined as the ratio of the average bootstrap variance to the empirical variance.

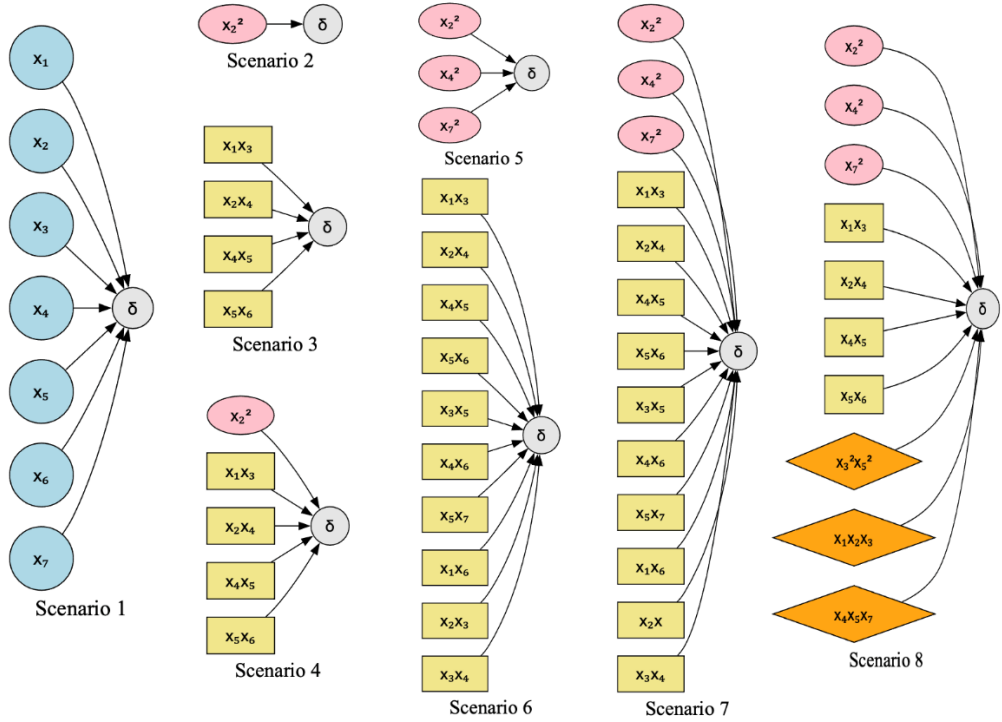

**Figure 1**: Functional forms of the true propensity models: Scenario 1 ($I_0Q_0$) linear; 2-4 ($I_0Q_1$, $I_1Q_0$, $I_1Q_1$) slightly, 5-7 ($I_0Q_2$, $I_2Q_0$, $I_2Q_2$) moderately, and 8($I_3Q_3$) severely nonlinear and/or nonadditive

## 4.3 Tuning Procedure

We tuned the GBM hyperparameters within each simulation run. Specifically, three key hyperparameters—learning rate (shrinkage), number of boosting iterations (trees), and maximum tree depth (interaction depth)—were optimized in every simulation replication, as these factors are known to influence GBM performance most. For each simulation run, the combination of hyperparameters that minimized covariate imbalance, as measured by the average standardized mean difference (ASMD), was selected and used for analysis consistently across all 1000 simulation runs. This approach ensures optimal covariate balance in each simulated dataset.

## 4.4 Results

Figure 2 illustrates the relative bias of the four pseudo-weighted estimators of the $FP$ quantity $\mu$. $_1$**PS** and $_2$**PS**, which incorporate main effects and two-way interactions in their logistic regression models, achieved only partial bias reduction. In simpler scenarios (1-4), these methods overall performed better than GBM-based methods and approximately unbiased, as their model structure was closer to the true propensity score (PS) model. However, as the true PS model became more complex, the bias in $_1$**PS** and $_2$**PS** increased notably. $_1$**PS** was especially prone to model misspecification, showing highly biased estimates in settings with moderate to high nonlinearity (e.g., 5 and 8). In such cases, estimates from $_1$**PS** fluctuated widely, with the absolute relative bias ranging from roughly 20% to 40%, indicating substantial instability under misspecification models. In contrast, although $_2$**PS** is also a parametric method and subject to misspecification, it benefits from a **two-step structure** that leverages the balancing score to mitigate bias. This design makes $_2$**PS** more stable than $_1$**PS** across complex scenarios, offering more consistent bias reduction even when the PS model is mis-specified.

Note that Scenario 6 –which includes main effects and moderate non-additivity with pairwise interactions, aligning with the true model for the traditional PS methods $_1$**PS** and $_2$**PS**— shows comparable performance across all four methods. When the PS model includes only quadratic terms or both quadratic terms and interactions (Scenarios 5, 7, 8), GBM-based estimators outperform the traditional methods, with **Boost$_2$PS** consistently achieving the lowest absolute relative bias, especially in Scenario 8, where the PS model is most complex involving nonlinearity and higher-order interactions. This highlights the effectiveness of its two-step design and nonparametric in correcting for bias, especially when the true PS model deviates substantially from simple forms.

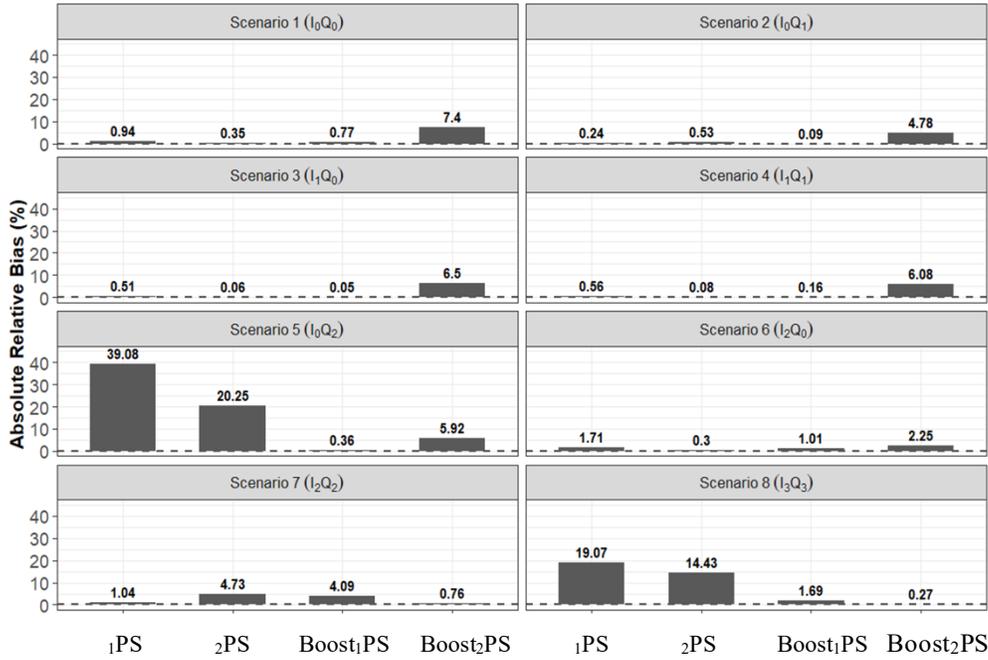

Figure 2. Absolute relative bias (%) of the traditional PS methods ($_1$PS, $_2$PS) and the GBM-based methods (Boost$_1$PS, Boost$_2$PS) across scenarios with varying degrees of nonlinearity ($Q_0$-$Q_3$) and non-additivity ($I_0$-$I_3$).

Table 1 presents the empirical variances across the eight scenarios for the four methods. Boost$_1$PS reduces variance relative to $_1$PS, in complex scenarios involving both nonlinear and nonadditive terms (e.g., Scenarios 7-8 of $I_2Q_2$ and $I_3Q_3$). However, the single stage methods ($_1$PS and Boost$_1$PS) consistently show higher empirical variances compared to the two-stage methods ($_2$PS and Boost$_2$PS). This difference is most pronounced in Scenario 8, where the variances of $_1$PS and Boost$_1$PS peak at 13.92 and 4.75, respectively, compared to 2.45 and 1.96 for $_2$PS and Boost$_2$PS. Across all scenarios, the empirical variance of $_2$PS remains moderate and stable, while Boost$_2$PS shows the most stable and lowest variances, often outperforming $_2$PS. These results suggest that the boosting procedures, especially Boost$_2$PS, effectively reduce variance under complex data structures. In line with these results, the mean squared error (MSE) of Boost$_2$PS is consistently among the lowest or comparable to the best-performing method, across all scenarios.

We also assessed the performance of the bootstrap variance estimators using the variance ratio (i.e., bootstrap variance divided by empirical variance), as presented in Table 4. Across all scenarios, the bootstrap variance generally overestimates the empirical variance, particularly for the GBM-based methods (Boost$_1$PS and Boost$_2$PS), with variance ratios ranging from approximately 1.15 to 1.27. These results suggest that the bootstrap procedure produces conservative variance estimates for the GBM-based estimators.

Table 1. Empirical variance, mean squared error and variance ratio of the population means estimates using the four methods

| Scenario | Empirical Variance ($\times 10^4$) | | | | Mean Squared Error ($\times 10^4$) | | | | Variance Ratio = Bootstrap Variance / Empirical Variance | | | |
|---|---|---|---|---|---|---|---|---|---|---|---|---|
| | $_1$PS | $_2$PS | Boost$_1$PS | Boost$_2$PS | $_1$PS | $_2$PS | Boost$_1$PS | Boost$_2$PS | $_1$PS | $_2$PS | Boost$_1$PS | Boost$_2$PS |
| $I_0Q_0$ | 2.18 | 2.02 | 3.85 | 2.04 | 2.20 | 2.03 | 3.86 | 3.69 | 1.03 | 1.01 | 1.23 | 1.25 |
| $I_0Q_1$ | 2.67 | 2.11 | 3.95 | 1.88 | 2.67 | 2.12 | 3.95 | 2.56 | 0.99 | 0.99 | 1.19 | 1.15 |
| $I_1Q_0$ | 2.42 | 2.19 | 3.41 | 1.82 | 2.43 | 2.19 | 3.41 | 3.09 | 1.03 | 1.00 | 1.23 | 1.29 |
| $I_1Q_1$ | 2.64 | 2.15 | 3.57 | 1.89 | 2.65 | 2.15 | 3.57 | 3.00 | 1.11 | 1.05 | 1.27 | 1.27 |
| $I_0Q_2$ | 4.99 | 2.23 | 5.22 | 2.61 | 50.87 | 14.54 | 5.22 | 3.66 | 1.24 | 1.09 | 1.21 | 1.15 |
| $I_2Q_0$ | 3.02 | 2.39 | 3.63 | 1.58 | 3.10 | 2.39 | 3.66 | 1.73 | 1.10 | 1.03 | 1.22 | 1.20 |
| $I_2Q_2$ | 6.38 | 2.68 | 4.16 | 1.92 | 6.41 | 3.35 | 4.67 | 1.94 | 1.17 | 1.06 | 1.22 | 1.15 |
| $I_3Q_3$ | 13.92 | 2.45 | 4.75 | 1.96 | 24.85 | 8.70 | 4.84 | 1.97 | 1.34 | 1.12 | 1.24 | 1.23 |

In summary, **Boost$_2$PS** demonstrates robust performance in both bias reduction and stability, consistently outperforming or matching alternative methods across a range of simple to complex propensity score model structures. Its flexibility and ability to accommodate model misspecification make it a strong candidate for pseudo-weighting nonprobability samples in settings with varying degrees of model complexity.

## 5. Real World Example

For illustration, we use the real-world data example in Wang et al. (2021). We evaluated multiple health outcomes over a 15-year period among U.S. adults, including all-cause mortality, cancer-related mortality, diabetes-related mortality, and heart disease-related mortality. This analysis involves the use of the adult household interview portion of the Third National Health and Nutrition Examination Survey (NHANES III), which was conducted from 1988 to 1994, with the sample size $n_c = 20,050$. We ignored all complex design features of NHANES III and approach it as a nonprobability sample. The coefficient of variation of the sample weights is 125%, indicating highly variable probabilities of sample selection, making the unweighted sample less representative and suggesting significant selection bias.

To facilitate comparability with an external survey, we treated NHANES III as a single cross-sectional sample and used data from the 1994 U.S. National Health Interview Survey (NHIS) supplement as the reference probability-based survey sample, which was specifically from respondents to the supplement that monitors progress toward the Healthy People Year 2000 objectives. The analysis included adults aged 18+, with a sample size of $n_s$ =19,738. The 1994 NHIS employed a multistage stratified cluster sampling design, consisting of 125 strata and 248 pseudo–primary sampling units (PSUs). To estimate variance, we collapsed strata with only one PSU with the next nearest stratum.

Using NHANES III as the "nonprobability sample" offers advantages for evaluating the effectiveness of propensity weighting methods. Both the NHANES III sample and the reference survey (NHIS) target similar populations, use the same data collection mode, and similar questionnaires. Moreover, both surveys have been linked to the National Death Index (NDI) for mortality follow-up of the respondents. These similarities increase the likelihood that the pseudo-weighted NHANES III sample can accurately represent the target population, thus can be helpful to evaluate the performance of propensity weighting methods using the sample-weighted NHIS estimates as the benchmark.

The propensity model includes variables of common demographic characteristics (age, sex race/ethnicity, region, and marital status), socioeconomic status (education level, poverty, and household income), tobacco usage (smoking status, and chewing tobacco), health variables (body mass index and self-reported health status), and a quadratic term for age. Table B in Appendix B presents the propensity logistic regression models fitted to the combined unweighted NHANES III and unweighted (or weighted) NHIS data.

We tuned our proposed GBM-based PS estimation methods using the NHANES III and 1994 NHIS datasets. As in the simulation study, the goal was to optimize the covariate balance between the non-probability sample (unweighted NHANES III) and the reference probability sample (NHIS) before constructing pseudo-weights. Given that the dimensionality and sample size of the real data are higher than the simulation scenarios, the search grid was expanded with a smaller shrinkage parameter (learning rate) and larger number of trees, because such configurations may enhance the performance despite longer computation times (McCaffrey et al. 2004). We also introduced a minimum

terminal-node--size parameter to prevent over splits and improve robustness.

Following the same tuning strategy used in the simulation, we performed a grid search over key GBM hyperparameters using the stacked samples of NHANES III and NHIS. The hyperparameters include learning rate (0.001, 0.0001), number of trees (1,000; 2,000; 5,000; 10,000), interaction depth (ranging from 4 to 10), and minimum number of observations for terminal nodes (5, 10, 15, 20). For each combination, we computed the ASMD to assess covariate balance. The configuration yielding the lowest ASMD was selected as the optimal set of hyperparameters. The optimal tuning parameters were 0.001 learning rate and 10,000 trees for both $Boost_1PS$ and $Boost_2PS$, with $Boost_1PS$ using depth 4 and 20 minimum observations per node, and $Boost_2PS$ using depth 8 and 10 minimum observations per node.

The propensity scores were estimated using GBM via the *gbm*() function from the *gbm* package in R, where the binary indicator of the survey source (= 1 for being from NHANES III, and 0 for being from NHIS) was modeled as the response variable and included abovementioned covariates. Balancing scores from the GBM model were extracted, representing each individual's estimated logit-likelihood of being in the NHANES sample given their covariates. The *ps*() function from the *twang* package was subsequently used to assess covariate balance based on the estimated scores, including the computation of standardized mean differences.

We first evaluated the performance of the $Boost_1PS$, and the $Boost_2PS$ method in balancing the covariate distributions in the NHANES III and the weighted NHIS sample. Figure 3 displays the distributions of the PS estimated by $Boost_2PS$ (on the logit scale) for the unweighted-, $Boost_1PS$-, and $Boost_2PS$-weighted NHANES III, compared to the sample-weighted NHIS. It can be observed that both $Boost_1PS$-weighted and $Boost_2PS$-weighted NHANES III samples closely approximate the distribution of the sample-weighted NHIS, with $Boost_2PS$ performing slightly better. We also compare the covariate balance between the sample-weighted NHIS and the NHANES III sample with or without pseudo-weights using the ASMD for each covariate that was included in the PS model (Table C in Appendix C). Consistent with Figure 3, the naive NHANES III sample shows large imbalances for many variables (e.g., Race, Education, Self-Reported Health), while all PS–adjusted methods substantially reduce ASMDs, especially $Boost_2PS$. GBM-based and two-step weighting approaches generally achieve better covariate balance than the single-step methods, particularly for the key sociodemographic and health-related variables. Overall, $Boost_2PS$ yields the closest alignment with the reference survey sample from NHIS, indicating the highest effectiveness in adjusting for differences.

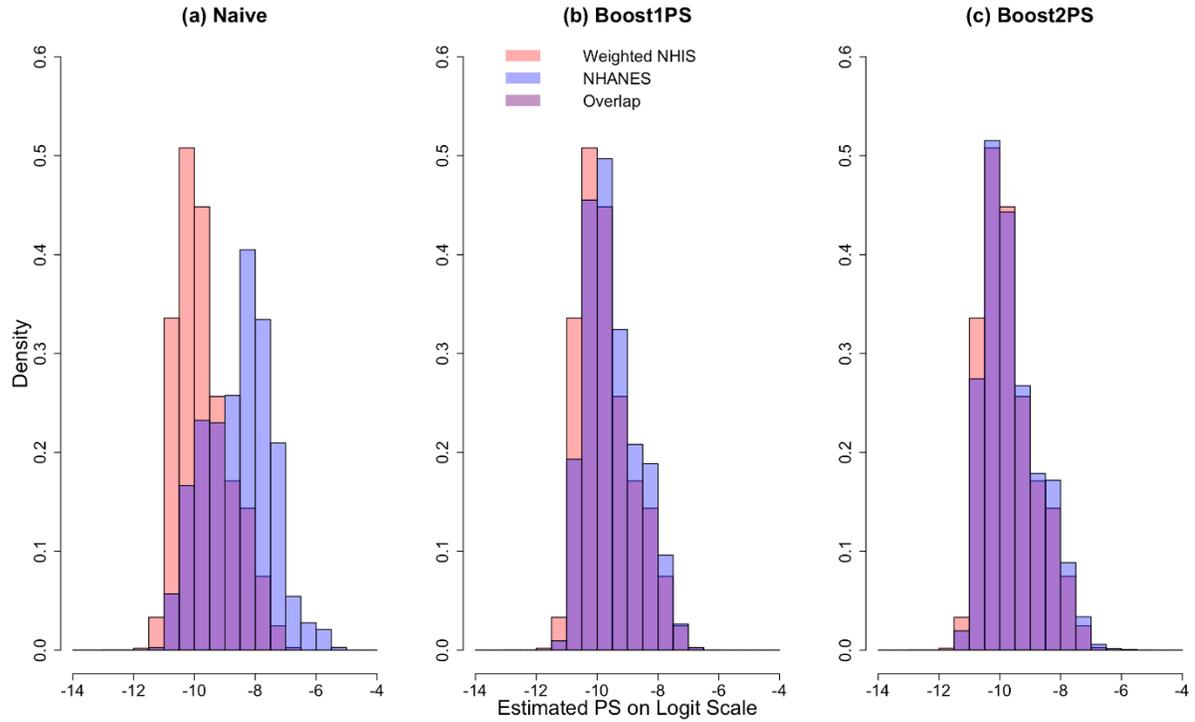

Figure 3. Distributions of the logit of the propensity scores (PS) estimated by Boost$_2$PS for (a) sample-weighted NHIS vs. naïve NHANES III; (b) sample-weighted NHIS vs. Boost$_1$PS-weighted NHANES III; and (c) sample-weighted NHIS vs. Boost$_2$PS-weighted NHANES III.

We then evaluated the performance of the Boost$_1$PS, and the Boost$_2$PS method in estimating finite population proportions of the mortality outcomes using relative bias(%RB), calculated against the sample-weighted NHIS estimates, $\%RB = \frac{\hat{\mu} - \mu_{\widehat{NHIS}}}{\mu_{\widehat{NHIS}}} \times 100$, treating the NHIS estimates as true, and bootstrap variances with the NHIS complex sampling designs considered (Wu and Rao, 1992).

The heatmap in Figure 4 displays the absolute relative bias (%) of the NHANES III estimates of the four mortality outcomes using the four pseudo-weighting methods or not, with the colors intensifying from light pink (lower bias) to dark red (higher bias). Unweighted NHANES III estimates show substantial bias. Boost$_2$PS consistently yields the smallest bias across most outcomes, particularly for survival and chronic conditions. Large biases are observed for diabetes-related mortality across all methods, though $Boost_2PS$ showing slight improvement over $_2PS$. Overall, $Boost_2PS$ outperforms other methods in bias reduction. As noted, Bias reduction for diabetes mortality was less effective compared to overall mortality, cancer mortality and heart disease, possibly due to omitted significant predictors of diabetes mortality and the limited ability of propensity models in capturing non-probability sample participation mechanisms of individuals at risk for diabetes mortality.

Figure 5 depicts a heatmap comparing the bootstrap standard error (SE) of different mortality outcomes using three methods: Naïve, $Boost_1PS$, and $Boost_2PS$. As with the simulation results, $Boost_2PS$ consistently outperforms $Boost_1PS$ with lower SEs, providing more reliable results across mortality outcomes.

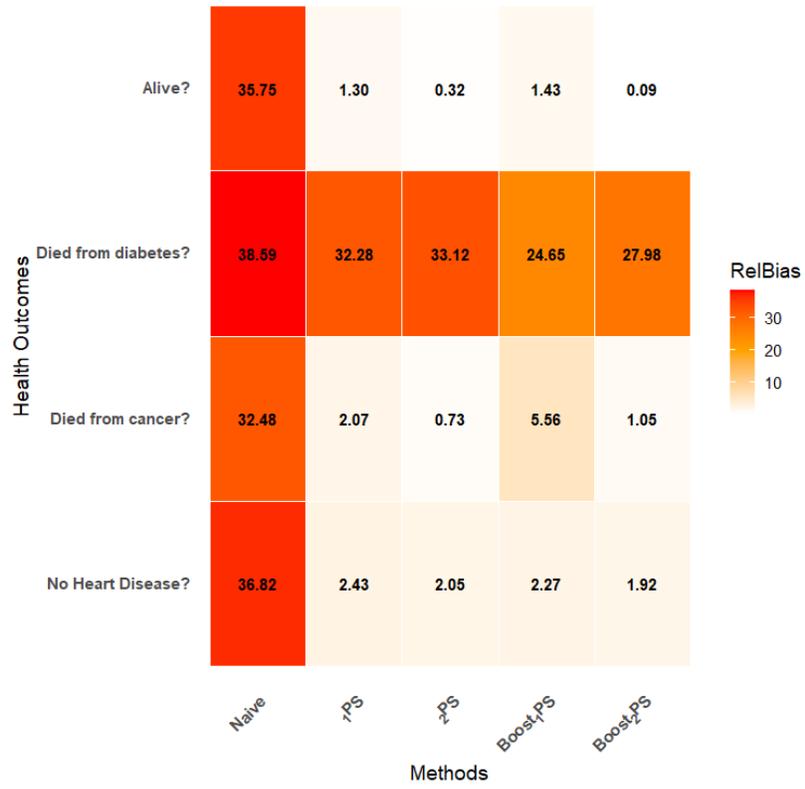

Figure 4: Absolute relative bias (%) of unweighted (naïve) and various pseudo-weighted NHANES III estimates against sample-weighted NHIS estimates.

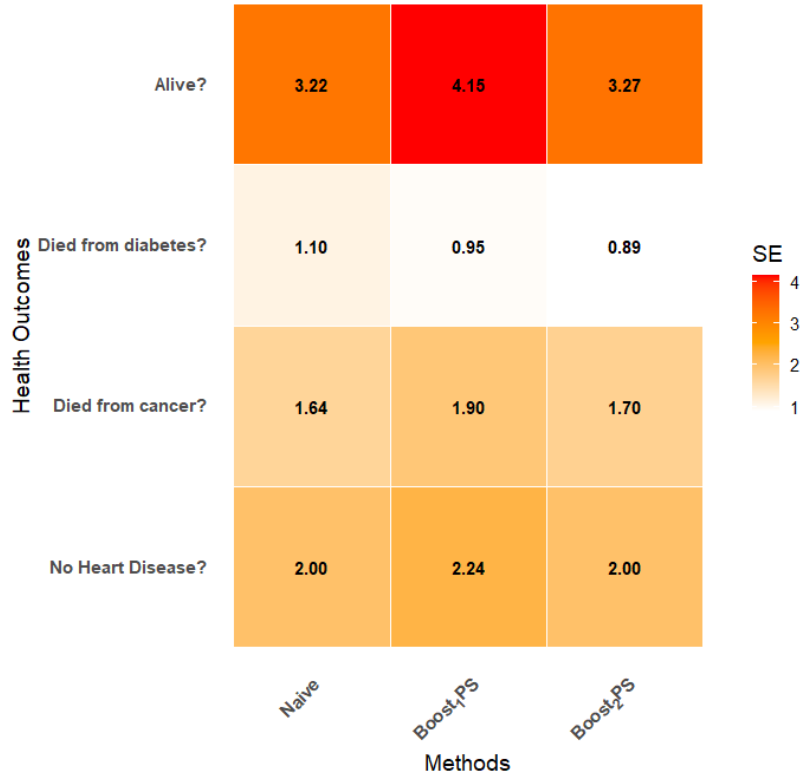

Figure 5: Bootstrap standard error ($\times 10^3$) of GBM-based NHANES III estimates compared to unweighted (naïve) NHANES III estimates.

## 6. Discussion

This paper developed Gradient-Boosted Pseudo-Weighting methods ($\text{Boost}_1\text{PS}$ and $\text{Boost}_2\text{PS}$) for population inference from nonprobability samples, aiming to mitigate selection bias. Unlike the traditional PS methods which are based on parametric models such as logistic regression, $\text{Boost}_2\text{PS}$ uses gradient boosting method (GBM) within a two-step ($_2\text{PS}$) pseudo-weighting framework. GBM offers greater flexibility in capturing complex, nonlinear relationships, which can improve covariate balance and reduce bias in population parameter estimation. We evaluated $\text{Boost}_2\text{PS}$ against other pseudo-weighting approaches, including $\text{Boost}_1\text{PS}$ and the traditional parametric methods, using both Monte Carlo simulations and real-world health survey data. Comparing to the benchmarks of the weighted NHIS estimates, Boost$_2$PS consistently reduced bias—especially under moderate to severe nonlinearity or non-additivity in the (self-)selection mechanism.

A key contribution of this work is the adaptation of loss functions in the gradient boosting algorithm for nonprobability sample inference, with and without incorporating survey weights. However, several limitations warrant attention. *First*, the second-step logistic adjustment in Boost$_2$PS assumes a reasonably good model fit, which should be further evaluated—potentially with design-based diagnostics from the probability sample. *Second*, overfitting remains a concern with boosting. Although GBM mitigates this risk through tree depth control, shrinkage (learning rate), and early stopping based on validation error (Friedman, 2001), overfitting can still occur, particularly with complex data or poorly tuned parameters. *Third*, the bootstrap variance ratio across methods ranged from 1.1 to 1.2, suggesting slightly conservative variance estimates. *Finally*, we focused on GBM in this paper because it requires fewer tuning parameters than some alternatives and is relatively robust to overfitting when tuned appropriately. Nevertheless, other boosting algorithms (Bhaduri et al, 2025) – such as AdaBoost (Freund & Schapire, 1997) or XGBoost (Chen & Guestrin, 2016) – may offer different bias-variance trade-offs and perform well on complex or large-scale datasets. While GBM remains a strong option for propensity score estimation, future research should explore comparisons with alternative boosting methods.

**Data Availability:** N/A

# Appendix A

We vary the coefficients of the $MOS^{(c)}$ in eight scenarios so that (1) the level of nonlinearity and non-additivity are different; and (2) the mean and variance of in the FP remain roughly constant across scenarios. As nonlinear and interaction terms emerge, smaller coefficients are required to prevent $MOS^{(c)}$ from inflating.

Scenario 1: A model with additivity and linearity

$$MOS^{(c)} = \exp\big(0.3(x_1 + x_2 + 1.5x_3 + 1.5x_4 - 0.8x_5 - 0.5x_6 + 0.7x_7)\big)$$

Scenario 2: A model with additivity and slight non-linearity

$$MOS^{(c)} = \exp\big(0.25(x_1 + x_2 + 1.5x_3 + 1.5x_4 - 0.8x_5 - 0.5x_6 + 0.7x_7 + x_2^2)\big)$$

Scenario 3: A model with slight non-additivity

$$MOS^{(c)} = \exp\big(0.27(x_1 + x_2 + 1.5x_3 + 1.5x_4 - 0.8x_5 - 0.5x_6 + 0.7x_7 + x_1x_3 + x_2x_4 + 1.5x_4x_5 - 0.8x_5x_6)\big)$$

Scenario 4: A model with slight non-additivity and slight non-linearity

$$MOS^{(c)} = \exp\big(0.25(x_1 + x_2 + 1.5x_3 + 1.5x_4 - 0.8x_5 - 0.5x_6 + 0.7x_7 + x_2^2 + x_1x_3 + x_2x_4 + 1.5x_4x_5 - 0.8x_5x_6)\big)$$

Scenario 5: A model with additivity and moderate non-linearity

$$MOS^{(c)} = \exp\big(0.25(x_1 + x_2 + 1.5x_3 + 1.5x_4 - 0.8x_5 - 0.5x_6 + 0.7x_7 + x_2^2 + 1.5x_4^2 + 0.7x_7^2)\big)$$

Scenario 6: A model with moderate non-additivity

$$MOS^{(c)} = \exp\big(0.22(x_1 + x_2 + 1.5x_3 + 1.5x_4 - 0.8x_5 - 0.5x_6 + 0.7x_7 + x_1x_3 + x_2x_4 + 1.5x_3x_5 + 1.5x_4x_6 - 0.8x_5x_7 + x_1x_6 + x_2x_3 + 1.5x_3x_4 + 1.5x_4x_5 - 0.8x_5x_6)\big)$$

Scenario 7: A model with moderate non-additivity and moderate non-linearity

$$MOS^{(c)} = \exp\big(0.17(x_1 + x_2 + 1.5x_3 + 1.5x_4 - 0.8x_5 - 0.5x_6 + 0.7x_7 + x_2^2 + 1.5x_4^2 + 0.7x_7^2 + x_1x_3 + x_2x_4 + 1.5x_3x_5 + 1.5x_4x_6 - 0.8x_5x_7 + x_1x_6 + x_2x_3 + 1.5x_3x_4 + 1.5x_4x_5 - 0.8x_5x_6)\big)$$

Scenario 8: A model with substantial non-additivity and substantial non-linearity

$$MOS^{(c)} = \exp\big(0.18(x_1 + x_2 + 1.5x_3 + 1.5x_4 - 0.8x_5 - 0.5x_6 + 0.7x_7 + x_2^2$$

$$+ 1.5x_4^2 + 0.7x_7^2 + x_1x_3 + x_2x_4 + 1.5x_4x_5 - 0.8x_5x_6$$

$$+ 1.5x_3^2x_5^2 + x_1x_2x_3 + x_4x_5x_7)\big)$$

## Appendix B

Table B: Estimated Coefficients from Propensity Score Models with and without considering NHIS Sample Weights

| | Unweighted NHIS | | | | Weighted NHIS | | | |
|---|---|---|---|---|---|---|---|---|
| | Estimate | Std. Error | t-value | P-value | Estimate | Std. Error | t-value | P-value |
| **(Intercept)** | 0.77 | 0.11 | 6.73 | < 0.01 | 0.44 | 0.14 | 3.09 | < 0.01 |
| Age (in years) | -0.08 | 0.00 | -21.49 | < 0.01 | -0.02 | 0.00 | -5.00 | < 0.01 |
| Age^2 | 0.00 | 0.00 | 24.45 | < 0.01 | 0.00 | 0.00 | 3.46 | < 0.01 |
| **Sex** (ref: male) | | | | | | | | |
| Sex: Female | -0.21 | 0.02 | -8.61 | < 0.01 | -0.06 | 0.03 | -1.88 | 0.06 |
| Education level | -0.15 | 0.01 | -15.35 | < 0.01 | -0.07 | 0.01 | -5.46 | < 0.01 |
| **Race/Ethnicity** (ref: NH-White) | | | | | | | | |
| Race: NH-Black | 1.41 | 0.03 | 44.39 | < 0.01 | -0.14 | 0.04 | -3.78 | < 0.01 |
| Race: Hispanic | 1.73 | 0.04 | 46.36 | < 0.01 | -0.18 | 0.05 | -3.51 | < 0.01 |
| Race: NH-Other | -0.09 | 0.07 | -1.25 | 0.21 | -0.19 | 0.09 | -2.20 | 0.03 |
| **Poverty** (ref: No) | | | | | | | | |
| Poverty: Yes | 0.15 | 0.04 | 3.82 | < 0.01 | -0.01 | 0.05 | -0.17 | 0.87 |
| Poverty: Unknown | 0.05 | 0.04 | 1.18 | 0.24 | 0.01 | 0.05 | 0.21 | 0.83 |
| Health Status | 0.25 | 0.01 | 21.79 | < 0.01 | 0.24 | 0.01 | 16.64 | < 0.01 |
| **Region** (ref: Northeast) | | | | | | | | |
| Region: Midwest | 0.09 | 0.04 | 2.52 | 0.01 | -0.06 | 0.04 | -1.41 | 0.16 |
| Region: South | 0.39 | 0.03 | 11.60 | < 0.01 | 0.03 | 0.04 | 0.80 | 0.43 |
| Region: West | 0.10 | 0.04 | 2.59 | < 0.01 | -0.04 | 0.05 | -0.85 | 0.39 |
| **Marital Status** (ref: married or living as married) | | | | | | | | |
| Marital Status: Single | -0.56 | 0.03 | -18.06 | < 0.01 | 0.00 | 0.04 | -0.11 | 0.91 |
| Marital Status: Previously married | -0.27 | 0.03 | -7.74 | < 0.01 | -0.01 | 0.05 | -0.30 | 0.77 |
| **Smoking** (ref: Non-smoker) | | | | | | | | |
| Smoking: Former smoker | 0.09 | 0.03 | 3.02 | < 0.01 | 0.08 | 0.04 | 2.32 | 0.02 |
| Smoking: Current smoker | 0.09 | 0.03 | 3.02 | < 0.01 | 0.12 | 0.04 | 3.25 | < 0.01 |
| Household Income | 0.07 | 0.01 | 9.44 | < 0.01 | 0.01 | 0.01 | 0.94 | 0.35 |
| **Chewing tobacco** (ref: No) | | | | | | | | |
| Chewing tobacco: Yes | -0.35 | 0.04 | -8.79 | < 0.01 | -0.31 | 0.05 | -6.08 | < 0.01 |
| **BMI** (ref: normal) | | | | | | | | |
| BMI: Under-weight | -0.07 | 0.07 | -1.00 | 0.32 | -0.13 | 0.08 | -1.51 | 0.13 |
| BMI: Over-weight | 0.01 | 0.03 | 0.41 | 0.68 | 0.00 | 0.03 | -0.02 | 0.98 |
| BMI: Obese | -0.05 | 0.03 | -1.48 | 0.14 | -0.07 | 0.04 | -1.62 | 0.11 |

# Appendix C

Table C: Standardized Mean Difference (SMD) Between Sample-Weighted NHIS and Pseudo-Weighted NHANES III

| Variable | Value | Weighted NHIS % | Unweighted NHANES III % | SMD | $_1$PS-Weighted NHANES III % | SMD | Boost$_1$PS-weighted NHANES III % | SMD | $_2$PS-weighted NHANES III % | SMD | Boost$_2$PS-weighted NHANES III % | SMD |
|---|---|---|---|---|---|---|---|---|---|---|---|---|
| | Sum of Pseudo-Weights | 186,924,967 | 20,050 | | 190,573,872 | | 172,225,925 | | 187,653,271 | | 180,737,145 | |
| | Value | % | % | SMD | % | SMD | % | SMD | % | SMD | % | SMD |
| Sex | Male | 47.7 | 46.9 | 0.017 | 44.7 | 0.061 | 45.9 | 0.038 | 46.1 | 0.034 | 47.4 | 0.007 |
| Age Category | 18-24 years | 13.4 | 15.8 | 0.072 | 11.3 | 0.061 | 12.9 | 0.015 | 12.2 | 0.036 | 12.3 | 0.032 |
| | 25-34 years | 43.6 | 35.4 | 0.169 | 42.2 | 0.030 | 41.1 | 0.051 | 43 | 0.013 | 44.8 | 0.025 |
| | 35-44 years | 26.6 | 22.6 | 0.093 | 27.7 | 0.024 | 27.6 | 0.023 | 27.5 | 0.020 | 25.4 | 0.029 |
| | 45-54 years | 5.1 | 6.3 | 0.050 | 5.3 | 0.010 | 5.8 | 0.028 | 5.1 | 0.000 | 5.6 | 0.019 |
| | 55-64 years | 4.6 | 6.4 | 0.076 | 5.4 | 0.037 | 5.3 | 0.030 | 5.1 | 0.022 | 5.1 | 0.022 |
| | 65+ years | 6.8 | 13.5 | 0.214 | 8.1 | 0.043 | 7.3 | 0.019 | 7.2 | 0.015 | 6.8 | 0.003 |
| Education Level | Less than high school /No GED | 7.6 | 23.9 | 0.452 | 8.7 | 0.031 | 8.9 | 0.036 | 9.7 | 0.056 | 8.2 | 0.017 |
| | High school graduate/GED | 11.6 | 18.6 | 0.197 | 11.9 | 0.009 | 12.9 | 0.036 | 12.8 | 0.034 | 11.4 | 0.004 |
| | Some college, no degree | 37.1 | 30.8 | 0.133 | 31.6 | 0.115 | 36.6 | 0.011 | 32.5 | 0.098 | 37.2 | 0.003 |
| | Associate's degree | 22.5 | 15.1 | 0.189 | 20.5 | 0.050 | 20.6 | 0.047 | 20.3 | 0.056 | 21.6 | 0.023 |
| | Bachelor's degree | 12.3 | 6.7 | 0.195 | 14.1 | 0.063 | 11.6 | 0.023 | 13.2 | 0.032 | 12.2 | 0.002 |
| | Graduate or professional degree | 8.9 | 4.9 | 0.159 | 13.1 | 0.162 | 9.4 | 0.016 | 11.6 | 0.103 | 9.3 | 0.012 |
| Poverty Status | No | 82.5 | 67.9 | 0.333 | 82.6 | 0.004 | 83.7 | 0.029 | 81.5 | 0.022 | 83.1 | 0.014 |
| | Yes | 10.7 | 21.4 | 0.283 | 9.4 | 0.035 | 10.6 | 0.004 | 10.4 | 0.008 | 10.6 | 0.003 |
| | Unknown | 6.8 | 10.7 | 0.137 | 7.9 | 0.040 | 5.7 | 0.038 | 8 | 0.044 | 6.3 | 0.018 |
| Self-Reported Health Status | Excellent health | 32.5 | 15.3 | 0.415 | 28.6 | 0.094 | 27.4 | 0.124 | 27.8 | 0.115 | 31 | 0.036 |
| | Very good health | 29.3 | 23.7 | 0.128 | 32.5 | 0.073 | 30.1 | 0.018 | 32.2 | 0.066 | 29.8 | 0.011 |
| | Good health | 25.8 | 35.9 | 0.220 | 28.1 | 0.049 | 29.3 | 0.075 | 28.7 | 0.063 | 26.1 | 0.007 |
| | Fair health | 9 | 19.9 | 0.311 | 9 | 0.001 | 10.2 | 0.036 | 9.4 | 0.013 | 9.8 | 0.025 |
| | Poor health | 3.4 | 5.2 | 0.085 | 1.8 | 0.075 | 3 | 0.019 | 1.9 | 0.071 | 3.2 | 0.009 |
| Geographic Region | Northeast | 20.6 | 14.6 | 0.158 | 23.2 | 0.068 | 18.9 | 0.047 | 20.9 | 0.006 | 19.9 | 0.020 |
| | Midwest | 25 | 19.2 | 0.139 | 23.9 | 0.027 | 24.8 | 0.006 | 24.7 | 0.008 | 25.1 | 0.003 |
| | South | 32.6 | 42.7 | 0.210 | 33.3 | 0.015 | 35.9 | 0.069 | 33.2 | 0.012 | 33.5 | 0.018 |
| | West | 21.7 | 23.5 | 0.041 | 19.5 | 0.053 | 20.5 | 0.031 | 21.3 | 0.011 | 21.5 | 0.006 |
| Race | Non-Hispanic White | 75.8 | 42.3 | 0.726 | 75.7 | 0.002 | 74.8 | 0.022 | 75.6 | 0.004 | 75.2 | 0.014 |
| | Non-Hispanic Black | 11.2 | 27.4 | 0.411 | 12.4 | 0.031 | 12.9 | 0.044 | 11.6 | 0.009 | 12 | 0.021 |
| | Hispanic | 9 | 28.9 | 0.532 | 7.8 | 0.033 | 9.3 | 0.008 | 8.9 | 0.002 | 9.3 | 0.008 |
| | Non-Hispanic Other | 4 | 1.5 | 0.166 | 4.1 | 0.005 | 3 | 0.064 | 3.9 | 0.007 | 3.5 | 0.031 |
| Marital Status | Married / Living with partner | 63.5 | 57.3 | 0.126 | 62.4 | 0.023 | 64.7 | 0.023 | 63.6 | 0.001 | 65.7 | 0.043 |
| | Widowed / Divorced / Separated | 17.5 | 22.1 | 0.107 | 20.1 | 0.062 | 17.6 | 0.003 | 18.5 | 0.024 | 16.8 | 0.015 |
| | Never married | 19 | 20.6 | 0.041 | 17.5 | 0.037 | 17.7 | 0.032 | 17.9 | 0.027 | 17.5 | 0.037 |
| BMI Category | [18.5,25) | 47.1 | 42.5 | 0.093 | 47.8 | 0.014 | 46.5 | 0.012 | 47.1 | 0.000 | 47 | 0.002 |
| | [0,18.5) | 3 | 2.9 | 0.005 | 2.8 | 0.015 | 2.7 | 0.016 | 2.9 | 0.009 | 2.8 | 0.015 |
| | [25,30) | 33.7 | 36 | 0.047 | 32.9 | 0.017 | 34.3 | 0.012 | 33.5 | 0.005 | 34.2 | 0.010 |

| | | | | | | | | | | | |
|---|---|---|---|---|---|---|---|---|---|---|---|
| | [30,81) | 16.1 | 18.6 | 0.065 | 16.5 | 0.009 | 16.4 | 0.008 | 16.5 | 0.010 | 16 | 0.004 |
| Family Income | Lower income | 29.8 | 48.4 | 0.378 | 29.6 | 0.004 | 31.1 | 0.025 | 30.3 | 0.008 | 29.4 | 0.009 |
| Smokeless Tobacco Use | No | 88.4 | 91.4 | 0.103 | 89.7 | 0.044 | 90.7 | 0.077 | 89.1 | 0.025 | 89.5 | 0.037 |
| Smoking Status | Never smoker | 50.3 | 51 | 0.015 | 52.1 | 0.035 | 48.7 | 0.033 | 50.7 | 0.007 | 49.7 | 0.012 |
| | Former smoker | 25.5 | 25 | 0.011 | 23.4 | 0.048 | 25.6 | 0.004 | 24.5 | 0.023 | 25.4 | 0.000 |
| | Current smoker | 24.2 | 24 | 0.006 | 24.5 | 0.008 | 25.7 | 0.034 | 24.9 | 0.015 | 24.8 | 0.014 |